\documentclass[utf8]{frontiersSCNS} 

\usepackage{url,hyperref,lineno,microtype,subcaption}
\usepackage[onehalfspacing]{setspace}

\nolinenumbers

\def\keyFont{\fontsize{8}{11}\helveticabold }
\def\firstAuthorLast{Song and Yang} 
\def\Authors{Xinyi Song\,$^{1}$ and Jun Yang\,$^{1,*}$}

\def\Address{$^{1}$Department of Atmospheric and Oceanic Sciences, School of Physics, Peking University, Beijing 100871, China}

\def\corrAuthor{Jun Yang}

\def\corrEmail{junyang@pku.edu.cn}

\begin{document}
\onecolumn
\firstpage{1}

\title[Asymmetry and Variability in Transmission Spectra]{Asymmetry and Variability in the Transmission Spectra of Tidally Locked Habitable Planets} 

\author[\firstAuthorLast ]{\Authors} 
\address{} 
\correspondance{} 

\extraAuth{}

\maketitle

\begin{abstract}

Spatial heterogeneity and temporal variability are general features in planetary weather and climate, due to the effects of planetary rotation, uneven stellar flux distribution,  fluid motion instability, etc. In this study, we investigate the asymmetry and variability in the transmission spectra of 1:1 spin--orbit tidally locked (or called synchronously rotating) planets around low-mass stars. We find that for rapidly rotating planets, the transit atmospheric thickness on the evening terminator (east of the substellar region) is significantly larger than that of the morning terminator (west of the substellar region). The asymmetry is mainly related to the spatial heterogeneity in ice clouds, as the contributions of liquid clouds and water vapor are smaller. The underlying mechanism is that there are always more ice clouds on the evening terminator, due to the combined effect of coupled Rossby--Kelvin waves and equatorial superrotation that advect vapor and clouds to the east, especially at high levels of the atmosphere. For slowly rotating planets, the asymmetry reverses (the morning terminator has a larger transmission depth than the evening terminator) but the magnitude is small or even negligible. For both rapidly and slowly rotating planets, there is strong variability in the transmission spectra. The asymmetry signal is nearly impossible to be observed by the James Webb Space Telescope (JWST), because the magnitude of the asymmetry (about 10 ppm) is smaller than the instrumental noise and the high variability further increases the challenge.

\tiny
 \keyFont{ \section{Keywords:} transit spectroscopy, exoplanet atmospheres, superrotation, asymmetry, varability} 
\end{abstract}

\section{Introduction} \label{sec:intro}

More than 4600 exoplanets have been discovered since the discovery of 51 Pegasi b in 1995 \citep{mayor1995jupiter}, with an ever-growing fraction of terrestrial planets that are smaller than Neptune and Uranus. Astronomers are now at the stage of characterizing atmospheric compositions of rocky planets. Transmission spectrum, emission spectrum, and reflection spectrum are three of the main methods used for atmospheric characterizations \citep{seager2000theoretical, hubbard2001theory,charbonneau2002detection, deming2017illusion,kaltenegger2017characterize,Kreidberg_2017,de2018atmospheric,grimm2018nature,schwieterman2018exoplanet}. In this study, we focus on the transmission spectrum. Stellar light is absorbed and scattered by atmospheric species when traveling through the planetary limb, therefore, at wavelengths where the molecules in the atmosphere have higher opacity, the planet will block more light from the star, having a higher relative transit depth.
The upcoming launch of telescopes such as JWST may enable us to detect transmission spectra with higher resolution, larger spectral coverage and longer observing duration, which will be a breakthrough in this area, especially for Earth-size planets.

Transmission spectroscopy is useful and effective in detecting atmospheric compositions of hot Jupiters, and even hot sub-Neptunes and super-Earths \citep{sing2011gran,nikolov2014hubble,wakeford2015transmission,benneke2019water,iyer2020influence,mikal2020transmission}. However, present telescopes are not able to resolve the atmospheric spectra of habitable terrestrial planets that are smaller in size and cooler in temperature. Theoretical analyses and numerical simulations showed that atmospheric CO$_2$ at 4.3\,$\mu$m on tidally locked planets is detectable by JWST, but not for other molecular signatures such as H$_2$O, because clouds (as well as haze) strongly mute the spectral features \citep{snellen2017detecting,lincowski2018evolved,fauchez2019impact,komacek2019atmospheric,lustig2019detectability,lacy2020jwst,pidhorodetska2020detectability,suissa2020dim}. Clouds cause a 10–100 times increase in the transit number required to detect H$_2$O using JWST \citep{komacek2019atmospheric}. Due to convection and large-scale circulation, the dayside of tidally locked habitable planets should be covered by clouds, especially over the substellar region \citep{yang2013stabilizing,yang2019ocean,haqq2018demarcating, fauchez2019impact, komacek2019atmospheric, suissa2020dim}.

In estimating the transmission spectra of tidally locked planets, previous studies \citep{fauchez2019impact,komacek2020clouds,suissa2020dim} calculated the average of the two terminators, assuming the differences between morning terminator and evening terminator are small or negligible. However, this assumption is groundless. Due to the single-direction rotation of the planet (clockwise or counterclockwise) and flow motion instability, asymmetry is unavoidable even for an aqua-planet with an ocean covering the whole surface. The atmospheric circulation on tidally locked planets is characterized by coupled Rossby-Kelvin waves and equatorial superrotation \citep{showman2010matsuno,showman2011equatorial,Kumar_2016,haqq2018demarcating}. Rossby-Kelvin waves and equatorial superrotation are able to advect heat, clouds, and species (such as H$_2$O) to the east of the substellar point, and can induce spatial asymmetry. We find that the asymmetry in the transmission spectra between the two terminators is significant and is comparable to the molecular signatures of H$_2$O in magnitude (see section~\ref{sec_results} below).

The feature of significant asymmetry in transmission spectra has already been found in the observations and simulations of hot exoplanets \citep{dobbs2012impact,line2016influence,von2016inferring,parmentier2018thermal,powell2019transit,ehrenreich2020nightside,lacy2020jwst,pluriel2020strong,kesseli2021confirmation}, but not for cool, habitable terrestrial planets. For example, an asymmetric feature in iron absorption of the hot Jupiter WASP-76b was confirmed by the High Accuracy Radial velocity Planet Searcher (HARPS) \citep{kesseli2021confirmation,ehrenreich2020nightside}. The asymmetry is always caused by nonuniform cloud coverage \citep{line2016influence}, which in turn is largely determined by thermal contrasts and atmospheric dynamics.

Another feature that was always ignored in the previous studies of terrestrial planets \citep{lincowski2018evolved,lustig2019detectability,fauchez2019impact,komacek2019atmospheric,lacy2020jwst,pidhorodetska2020detectability,suissa2020dim,suissa2020first} is the variability of the transmission spectra. On account of waves and fluid motion instabilities, there may be strong short-term variability in the atmosphere although there is neither seasonal cycle nor diurnal cycle on the 1:1 tidally locked planets. The variability is unimportant for the mean climate but it is critical for atmospheric characterizations. This is because a strong (weak) variability implies that more (less) observation time is required for proper resolving of the molecular signals or the asymmetry. Previous studies of brown dwarfs and sub-Neptunes (such as \citealp{artigau2009photometric,radigan2012large,apai2013hst,showman2013atmospheric,showman2019atmospheric,charnay2020formation}) have found that the cloud fraction at the terminators could be highly variable.

Recent model calculations by \cite{may2021water} shows that the variability of patchy ice cloud is present in the upper atmospheres of M-dwarf terrestrial planets, particularly along the limbs. However, their study still uses the same method as previous studies, assuming the differences between morning terminator and evening terminator are small or negligible, and calculated the average of the two terminators. Here in our study, we examine the asymmetry in the transmission spectra between morning terminator and evening terminator, and the variability of this asymmetry. Moreover, while \cite{may2021water} examine only one rapidly rotating planet TRAPPIST-1e, our study examines the results of both rapidly and slowly rotating planets. Note that in order to have habitable environments permitting the existence of liquid water, a planet has to be rather close to an M dwarf, such as 0.1 AU, since M dwarfs are relatively cool and dim. Such a small orbit would result in a significant gravity gradient along the planet’s diameter. The gravitational pull on the planet can lead to torques that force the planet to synchronize its rotation period with its orbital period. The time scale for the evolution of one habitable planet from asynchronous rotation to synchronous rotation depends on many factors, such as the orbital distance, the stellar mass, the energy dissipation rate, the rigidity of the planet, the initial planetary rotation rate, land-sea configuration, ocean bottom topography \citep{Kasting_1993,barnes2017tidal,pierrehumbert2019atmospheric}. Habitable planets around low-mass stars are more likely to be in synchronous rotation than Earth.

To calculate the asymmetry and variability between morning and evening terminators in transmission spectra of tidally locked terrestrial planets, we post-process 3D climatic outputs of the atmospheric general circulation model (AGCM) in \cite{zhang2020does} to calculate JWST observations of transmission spectra for morning and evening terminators, respectively. In section~\ref{sec_methods}, we describe the settings for the AGCM experiments, and how we post-process the AGCM results to obtain the transmission spectra and its variability. Then, we present the transmission spectra and estimate the number of transits required to detect this asymmetry signal in section~\ref{sec_results}. A summary is given in section~\ref{sec_summary}.


\section{Methods}\label{sec_methods}

\subsection{Simulation of the Climate}\label{sec_GCM}
We employed the three-dimensional (3D) atmospheric general circulation model (AGCM) ExoCAM in our climate simulations \citep{zhang2020does}. ExoCAM was developed based on the Community Atmosphere Model version 4 but with correlated-$k$ radiative transfer and had updated water vapor continuum absorption coefficients from HITRAN 2012 \citep{wolf2014controls,wolf2017assessing,wolf2017constraints}. In the experiments, planetary radius, gravity, and atmospheric conditions are the same as Earth, except we only include N$_2$ and H$_2$O. The atmosphere is coupled to a 50-m immobile, slab ocean with no oceanic heat transport and no continent. The horizontal resolution is 4$^{\circ}$ by 5$^{\circ}$ (46 grid points in latitude and 72 grid points in longitude). The model has 40 layers vertically, with surface pressure of 1~bar and model top of 1~hPa. The stellar flux is 1350 W\,m$^{-2}$, and the stellar spectra are from the BT\_Settl stellar model \citep{allard2007k}. The stellar temperature is 2600~K for the rapidly rotating planet and 3700 K for the slowly rotating planet.

Since the simulations presented in this paper are non-runaway planets, which means their cloud deck lies well below the original model top, there is no need to extend the layers of the atmospheric profiles \citep{suissa2020dim}. The results of two cases, with rotation periods of 6 and 60 Earth days, are shown in the main text, as they are the most representative ones. Results of other experiments with different surface pressures and stellar fluxes are discussed at the end of the paper. The experiments are set to be 1:1 tidally locked (rotation period = orbital period). Each of the experiments were run for tens of Earth years. We use the 365 daily outputs of the last year of the experiments to analyze the transmission spectra and their viability. 

Our experimental design is different from that of \cite{Kumar_2016, kumar2017habitable}, who modified the rotation period and the stellar flux simultaneously in their experiments, self-consistently considering the effect of the Coriolis force and stellar flux as a whole instead of two standalone factors. We use the same method as \cite{merlis2010atmospheric}, \cite{way2016venus}, \cite{noda2017circulation}, and \cite{Bin_2018}, with the rotation period fixed when varying the stellar flux or with the stellar flux fixed when varying the rotation period. This method allows us to know the separate effects of varying the rotation period and varying the stellar flux. The combined effect of varying the rotation period and the stellar flux needs future studies.


\subsection{Calculation of the Transmission Spectra}\label{sec_PSG}

\begin{figure}
\begin{center}
    \includegraphics[width=0.5\textwidth]{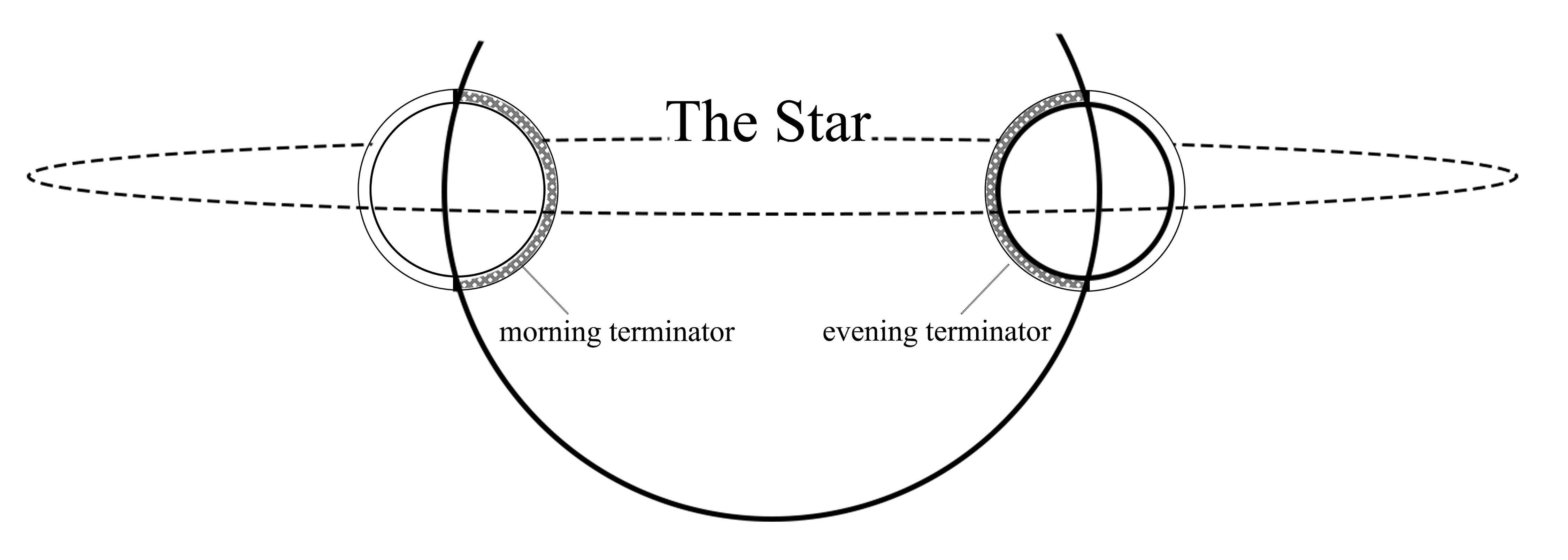}
    \caption{The orbital geometry of a transiting exoplanet system. The dashed line is the planetary orbit, the thick solid line is the edge of the star, and two phases of the planet during the ingress and egress are shown. The asymmetry between the transit depth of the morning terminator during the ingress and the transit depth of the evening terminator during the egress leads to a distorted light curve.}
    \label{fig_geometry}
\end{center}
\end{figure}

We use the Planetary Spectrum Generator\footnote{ \url{ https://psg.gsfc.nasa.gov}} (PSG, \citealp{villanueva2018planetary}) to calculate the  transmission spectra. The method used here is similar to that used in \cite{fauchez2019impact}, \cite{suissa2020dim} and \cite{komacek2020clouds}. They averaged transmission spectra of the two terminators, but we calculate transmission spectra for the morning terminator and evening terminator, respectively. In theory, the transmission spectra of the morning terminator can be obtained during the ingress, and the transmission spectra of the evening terminator can be obtained during the egress, as shown in Figure~\ref{fig_geometry}. In practice, the process for obtaining the separate morning and evening spectra is more complex; please refer to the paper by \cite{espinoza2021constraining}. What's more, unlike previous studies (such as \citealp{komacek2020clouds}) that use the climatology outputs averaged over the last 10 years of the simulation time, we use the daily output of the experiments, so our analyses are able to resolve the variability of the spectra. We use PSG to calculate the transmission spectrum for every model grid point along the terminator, and then average the spectra with equal weighting of each grid. 

We calculate the transmission spectra with a resolving power (R) of 300, from 0.6 to 5 $\mu$m, within the range of the Near Infrared Spectrograph instrument (NIRSpec) on JWST, and it has been shown that NIRSpec is the best instrument for JWST characterization of terrestrial exoplanets \citep{Batalha2018Strategies,lincowski2019observing,lustig2019detectability}. Since we do not include a noise floor and only use the PSG imager noise model in the calculations, the estimated results for the number of required transits to detect the spectral features should be considered as a lower limit. We set the output of PSG in the form of relative transit atmospheric thickness. The relative transit depth can be calculated by the following mathematical expression  \citep{winn2010exoplanet}: 
\begin{equation}
\delta \Delta=(\delta R/R_s)^2 + (2 \times R_p  \times \delta R) / R_s^2 \approx (2 \times R_p  \times \delta R) / R_s^2
\end{equation}
with $R_p$, $\delta R$, and $R_s$ the planet’s radius, transit atmospheric thickness, and the radius of the star, respectively. $\Delta$ is the transit depth, and $\delta \Delta$ is the relative transit depth.
Note that during the following calculations, we assume that the rapidly rotating case and the slowly rotating case have the same stellar radius as TRAPPIST-1 (0.12 solar radius), and the planet radius the same as TRAPPIST-1e (0.91 Earth's radius). The influence of stellar radius is further discussed in section~\ref{sec_summary}.

\begin{figure}
    \includegraphics[width=1.0\textwidth]{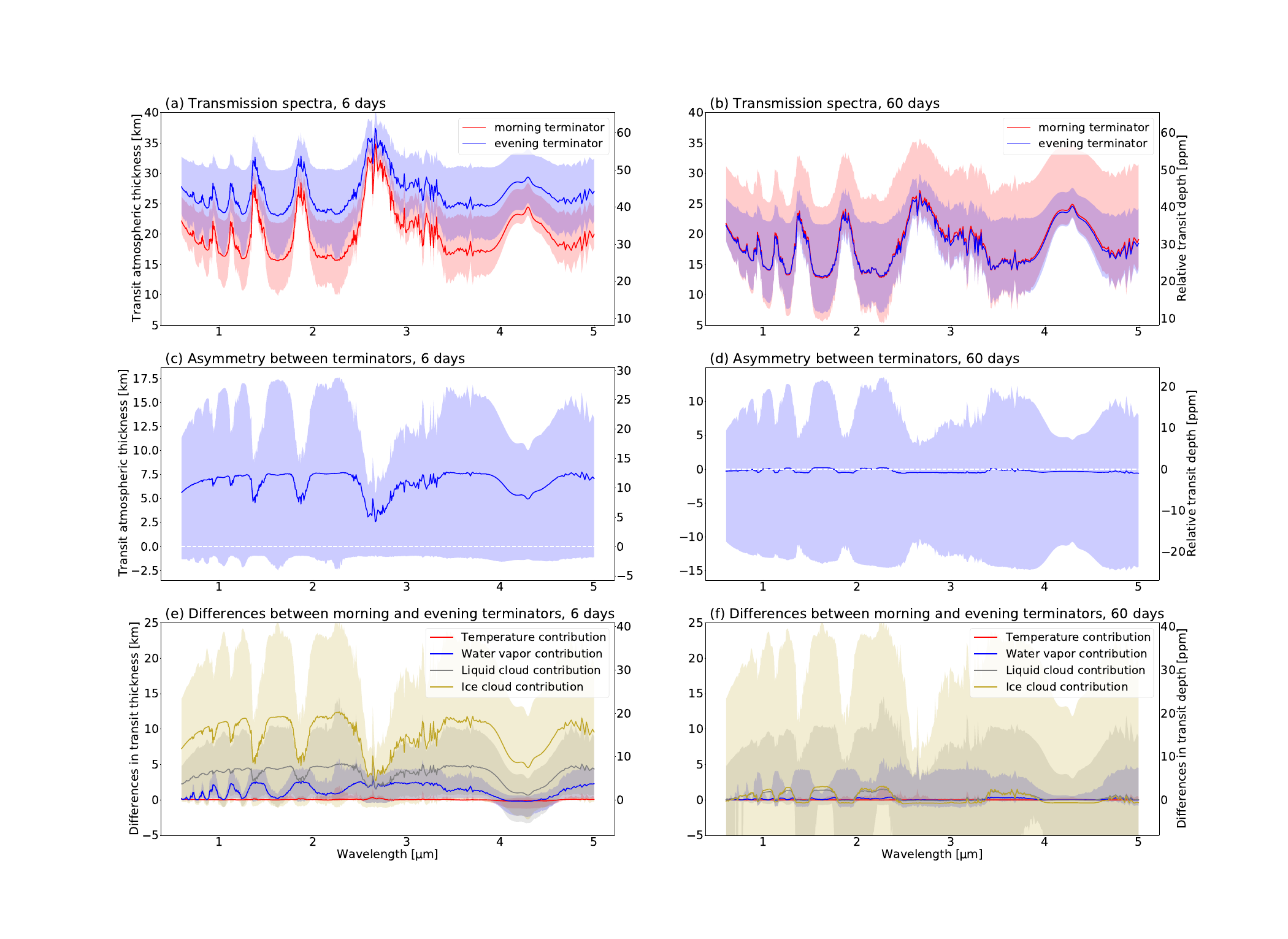}
    \caption{Transmission spectra and their variability of tidally locked planets, from 0.6 to 5 $\mu$m. The left y axis is the transit atmospheric thickness (km) and the right y axis is the relative transit depth (ppm). (a), (c) and (e): results for planets with a rotation period of 6 Earth days (= orbital period); (b), (d) and (f): for planets with a rotation period of 60 Earth days. In (a) and (b), the blue line is for the evening terminator and the red line is for the morning terminator, and the areas shaded with pale blue and pale red show their variability. In (c) and (d), the blue line is for the average value of the asymmetry (evening minus morning), and the areas shaded with pale blue show their variability. In (e) and (f), the contributions of air temperature, water vapor, liquid clouds, and ice clouds to the asymmetry (evening minus morning) are shown in red, blue, grey, and yellow, respectively, and their variability are shaded with the corresponding pale colors. The surface pressure is 1~bar, including only N$_2$ and H$_2$O. Note that the summary of the four lines in the lower panel (such as (e)) is not exactly equal to the difference between the blue line and the red line in the upper panel (such as (a)); this is mainly due to the overlaps among water vapor, liquid clouds, and ice clouds.}
    \label{fig_spectra}
\end{figure}


\section{Results} \label{sec_results}

\subsection{Asymmetry in the Transmission Spectra}\label{sec_asymmetry}

Figure~\ref{fig_spectra}(a) shows the transmission spectra for one tidally locked aqua-planet with a rotation period of 6 Earth days. It is clear that the transit atmospheric thickness of the morning terminator (west of the substellar point, red line) is in general smaller than that of the evening terminator (east of the substellar point, blue line) in every wavelength. The magnitude of the differences between these two lines is in the order of 10~ppm (Figure~\ref{fig_spectra}(c)). This magnitude is comparable to the molecular absorption signals of H$_2$O, such as the signal at 1.4 $\mu$m. In this figure, all of the absorption features are due to water vapor and clouds, except for the 4.3 $\mu$m feature caused by N$_2$--N$_2$ collision-induced absorption.

To determine which factor is the main cause of the asymmetry between morning and evening terminators, we calculated the contributions of air temperature, water vapor, liquid clouds, and ice clouds, respectively, as shown in Figure~\ref{fig_spectra}(e). When calculating the contribution of liquid clouds, for example, we set temperature, water vapor and ice clouds with the same values on both morning and evening terminators. It turns out that ice clouds contribute the most to the differences, while the contributions of the asymmetry in air temperature, water vapor, and liquid clouds are much smaller. This is due to the fact that ice clouds are in the high-altitude levels where the optical thicknesses of water vapor and liquid clouds are relatively smaller, thereby they have stronger influence on the transit depth.

\begin{figure}
\begin{center}
    \includegraphics[width=1.0\textwidth]{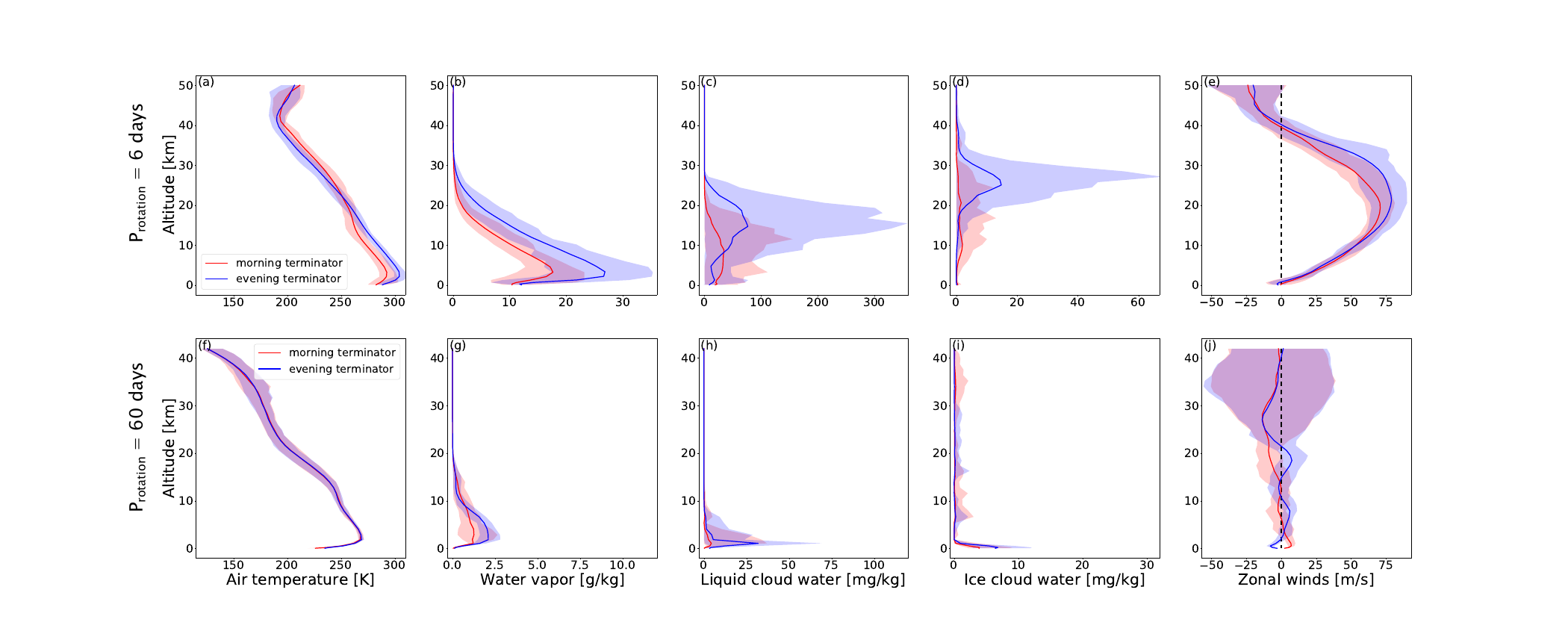}
    \caption{Asymmetry and variability in climatic variables. Profiles of air temperature (a \& f), water vapor concentration (b \& g), liquid cloud water (c \& h), ice cloud water (d \& i), and zonal (west--to--east) winds (e \& j) at the terminators. Red color is for the morning terminator and blue color is for the evening terminator. Lines are for the mean value and color shadings are for the variability. The rotation period (= orbital period) is 6 Earth days for the upper panels, and is 60 Earth days for the lower panels. The dashed lines in (e) and (j) are zero lines.}
    \label{fig_profiles}
\end{center}
\end{figure}

\begin{figure}
\begin{center}
    \includegraphics[width=1.00\textwidth]{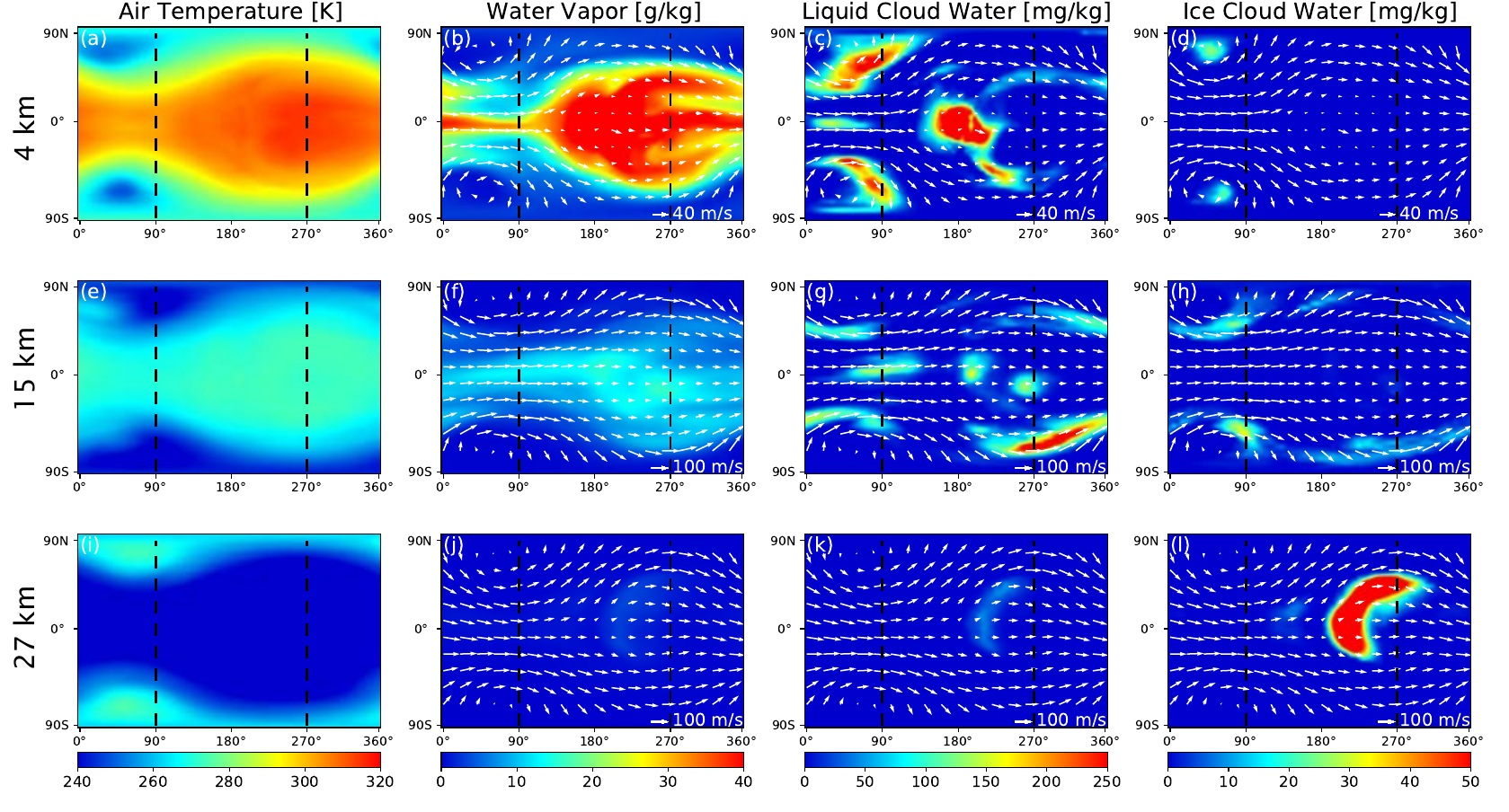}
    \caption{Spatial patterns of surface temperature (the first column), water vapor concentration (the 2nd column), liquid cloud water concentration (the 3rd column), and ice cloud water concentration (the 4th column) in one time snapshot at three different levels (4, 15, and 27~km), for the rapidly rotating planet of 6 Earth days. Each row represents one level. The arrows are the winds at the corresponding levels with a reference length of 40 m\,s$^{-1}$ for the level of 4~km and 100 m\,s$^{-1}$ for the levels of 15 and 27~km. The substellar point is located at the center of each panel, and the vertical dashed lines are for the morning terminator (90$^\circ$) and the evening terminator (270$^\circ$).}
    \label{fig_spatial_pattern1}
\end{center}
\end{figure}

Water vapor concentration at the evening terminator is greater than that at the morning terminator in every layer (Figure~\ref{fig_profiles}(b)). This is caused by atmospheric superrotation and coupled Kelvin-Rossby waves (excited from the uneven distribution of stellar flux; \cite{showman2010matsuno,showman2011equatorial,tsai2014three,hammond2018wave,pierrehumbert2019atmospheric,wang2021phase}), which advects low-concentration water vapor from the nightside to the west terminator and meanwhile advects high-concentration water vapor from the substellar region to the east of the substellar point, as shown in Figure~\ref{fig_spatial_pattern1}(b, f, \& j).

The vertical distributions of liquid cloud water path and ice cloud water path (Figure~\ref{fig_profiles}(c) and (d)) are more complex. For ice clouds, the concentration below the level of $\approx$20 km at the morning terminator is larger than that at the evening terminator, but above that level, the contrast reverses and the evening terminator has more ice clouds than the morning terminator. The liquid cloud concentration distribution is similar to that of ice cloud, but for liquid clouds the reverse occurs at a lower level of $\approx$10 km.

\begin{figure}
\begin{center}
    \includegraphics[width=1.00\textwidth]{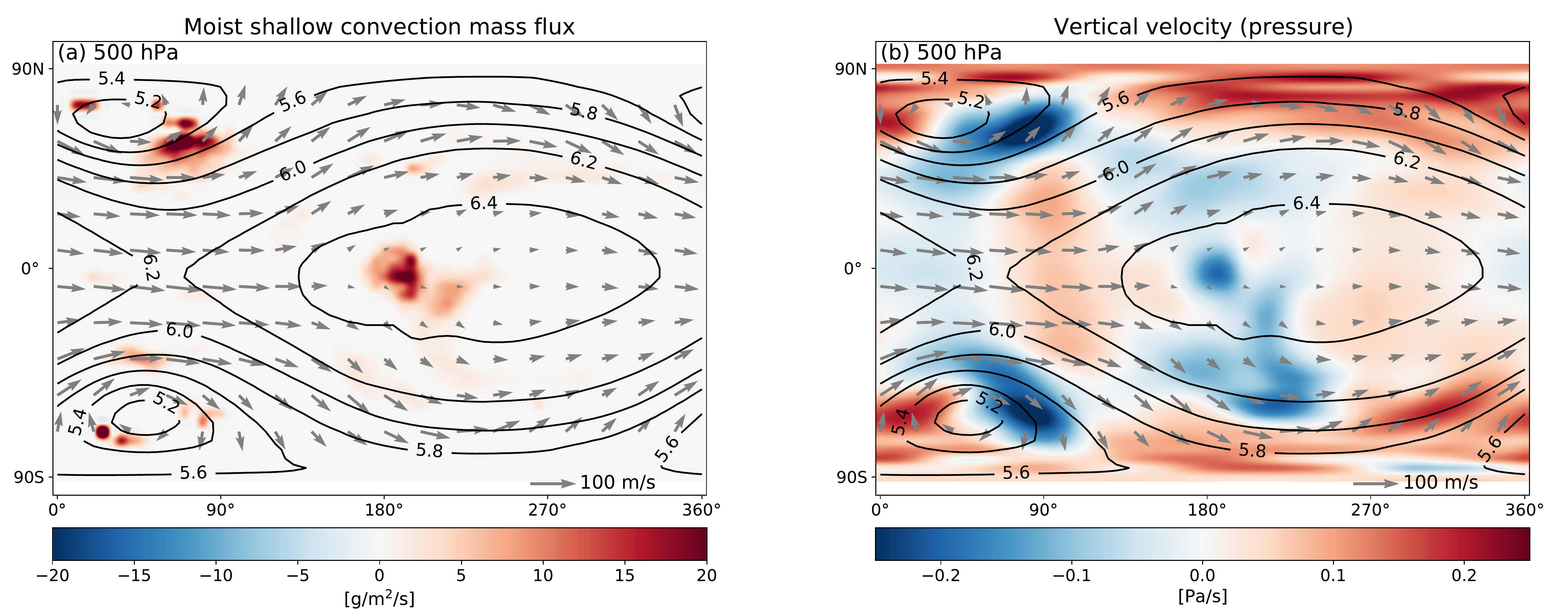}
    \caption{Moist shallow convection mass flux (g\,m$^{-2}$\,s$^{-1}$, color shading in (a)), horizontal winds at 500 hPa (m\,s$^{-1}$, vectors in both (a) and (b)), and vertical velocity at 500 hPa (Pa\,s$^{-1}$, color shading in (b)), in the rapidly rotating case of 6 Earth days. Red color in (b) stands for descending motion and blue color in (b) stands for upward motion. Note that besides the strong convection at the substellar region, there is also shallow convection at the two cyclones near the morning terminator (90$^\circ$).}
    \label{fig_convection}
\end{center}
\end{figure}

Detailed analyses find that clouds (both liquid and ice) at the morning terminator mainly form at the local region whereas clouds at the evening terminator are mainly advected from the substellar region by horizontal winds (Figure~\ref{fig_spatial_pattern1}).
There are two key regions where clouds form: one well-known region is the substellar area and the other barely-known one is the the cyclone region (characterized by low pressure, anticlockwise winds in the northern hemisphere and clockwise winds in the southern hemisphere, and upwelling motion) near the morning terminator, as shown in Figure~\ref{fig_convection}. Therefore, at low levels of the atmosphere, the morning terminator has more liquid and ice clouds than the evening terminator. At high levels (such as $\geq$20 km for ice clouds), the morning terminator has less ice clouds than the evening terminator; this is because the zonal westerly winds transport clouds from the substellar region to the east terminator (Figure~\ref{fig_spatial_pattern1}(l)). Note that the zonal winds are high-altitude (above $\approx$10~km) enhanced, so the eastward transport of clouds to the evening terminator is not significant below the level of $\approx$10~km for liquid clouds and below the level of $\approx$20~km for ice clouds.

For a slowly rotating planet of 60 Earth days, the asymmetry is much smaller than that of the rapidly rotating case (Figure~\ref{fig_spectra}(d) versus (c)). This is reasonable because as the rotation period increases, the baroclinic instability and the interaction between waves and mean flow become weaker \citep{vallis2017atmospheric}, so that the climatic variables (such as air temperature, water vapor, liquid clouds, and ice clouds shown in Figure~\ref{fig_profiles}(f, g, h, \& i)) are much more symmetric between the two terminators. Imagine a planet with no rotation and with uniform  boundary conditions (such as the aqua-planet employed here), the asymmetry may be close to zero.

Interestingly, the transit atmospheric thickness of the morning terminator is slightly larger than that of the evening terminator in the slowly rotating case; it is opposite to the result of the rapidly rotating case. This is due to the fact that the strength of atmospheric superrotation is much weaker in the slowly rotating case than in the rapidly rotating case (Figure~\ref{fig_profiles}(j) versus~\ref{fig_profiles}(e), same as that found in  \cite{Kumar_2016,kumar2017habitable,haqq2018demarcating}). Thus, horizontal advection has nearly no effect on the cloud concentration at the east (evening) terminator, whereas the west (morning) terminator, where clouds form locally, can have more clouds (such as at the levels above 30~km, shown in Figure~\ref{fig_profiles}(i)) and a larger transit depth (see Figure~\ref{fig_spectra}(b)) than the east terminator.

\begin{figure}
\begin{center}
    \includegraphics[width=1.00\textwidth]{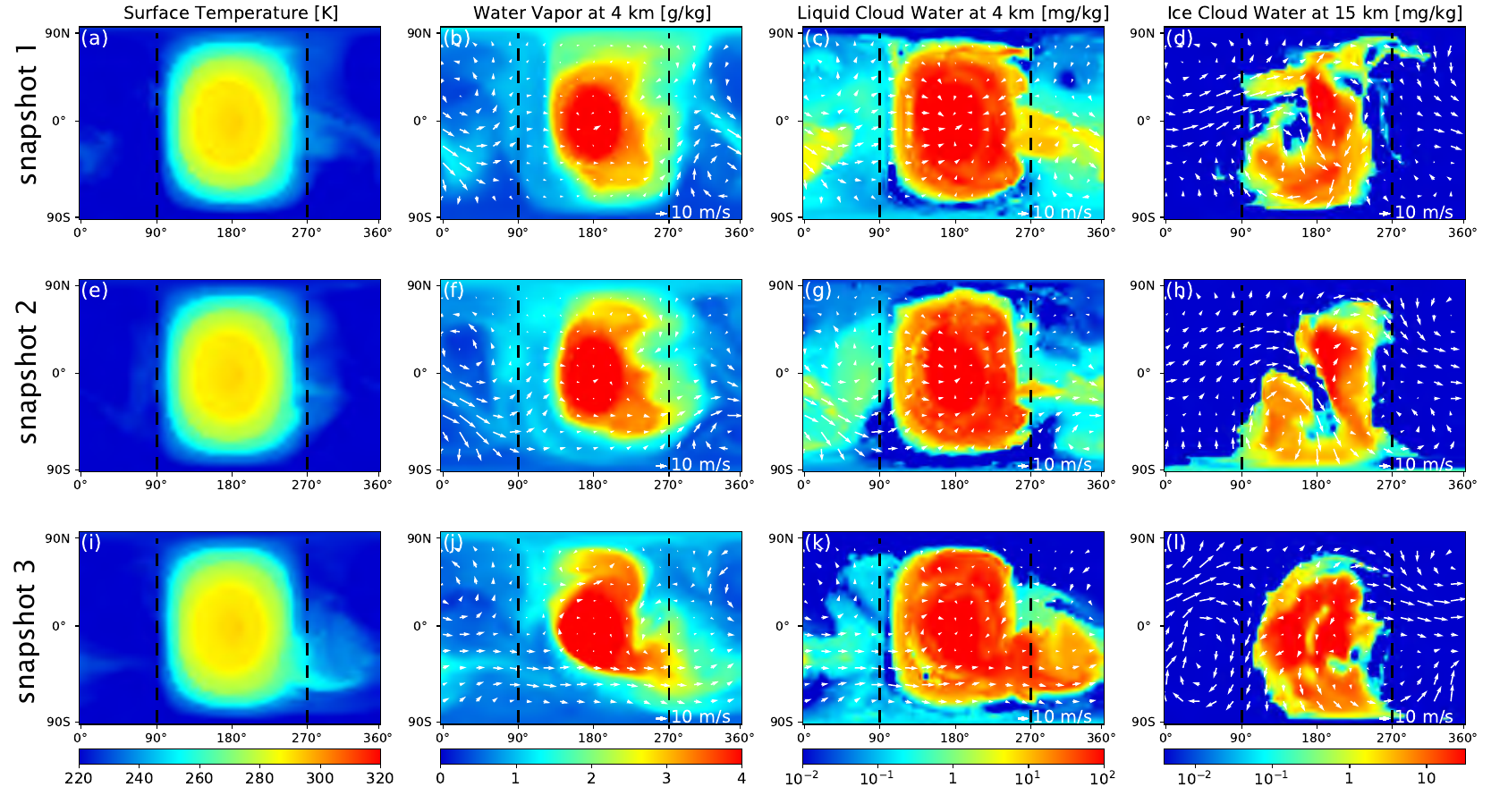}
    \caption{Variability of the system: surface temperature (the first column), water vapor concentration at 4~km (the 2nd column), liquid cloud water concentration at 4~km (the 3rd column), and ice cloud water concentration at 15~km (the 4th column) in three different time snapshots, for the slowly rotating planet of 60 Earth days. The reference vector is 10~m\,s$^{-1}$. Note that the color bars in the three right columns are different from those in Figure~\ref{fig_spatial_pattern1}. On this planet, the variability is still strong although the asymmetry between the morning and evening terminators is smaller than that on rapidly rotating planets in the transmission spectra (see Figure~\ref{fig_spectra}).}
    \label{fig_spatial_pattern2}
\end{center}
\end{figure}

\begin{figure}
\begin{center}
    \includegraphics[width=0.95\textwidth]{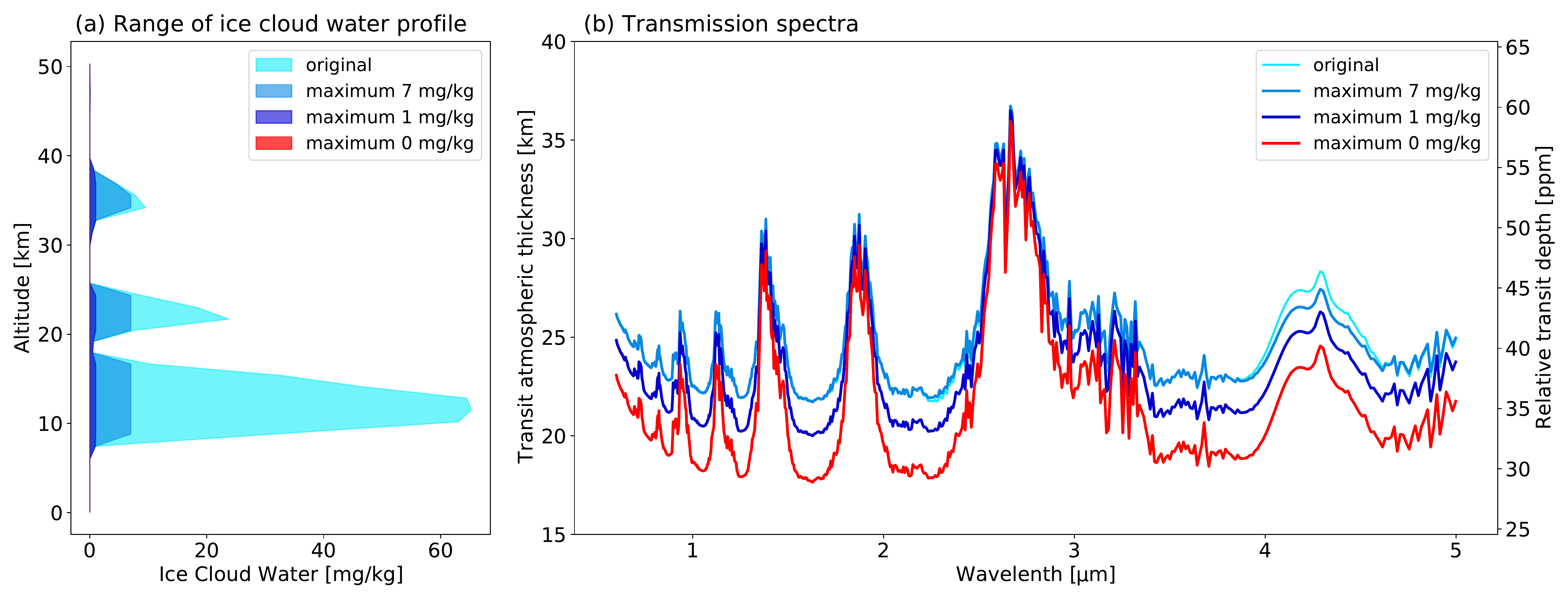}
    \caption{The saturation effect of ice cloud water on transmission thickness in the rapidly rotating case of 6 Earth days. The shadings in (a) show the range of ice cloud water content along the morning terminator. We artificially set the maximum value of ice cloud water to 0, 1, and 7 mg\,kg$^{-1}$ in (a), and their corresponding transit atmospheric thicknesses are shown in (b). When the maximum value of ice cloud water is set to 7 mg\,kg$^{-1}$, the transit atmospheric thickness is almost the same as the original, and we call this phenomenon ``the saturation effect".}
    \label{fig_along-lat}
\end{center}
\end{figure}



\subsection{Strong Variability in the Transmission Spectra}\label{sec_variability}

Besides the asymmetry, there is also strong variability in the transmission spectra, as the color shading showed in Figure~\ref{fig_spectra}. The amplitude of the variability is in the order of 20 ppm, larger than the asymmetry and H$_2$O signals. The variability in the transmission spectra is caused by the strong variability in water vapor, liquid clouds, and ice clouds, especially the latter, as shown in Figures~\ref{fig_profiles} and \ref{fig_spatial_pattern2}. For slowly rotating planets, the variability in the transmission spectra is comparable to that of the rapidly rotating planets, because the vortexes on slowly rotating planets move more freely (Figure \ref{fig_spatial_pattern2}), while those in the rapidly rotating case are confined to the region near the morning terminator. Although the spatial pattern of the surface temperature is relatively stable (left panels in Figure~\ref{fig_spatial_pattern2}), all of the waves, barotropic instability, baroclinic instability, and the interaction between the waves and mean flow are able to induce oscillations and randomness in the system. We tried Fourier analysis on the time series of transit atmospheric thickness and climate variables, but we didn't find any notable period. The strong variability and irregularity make the asymmetry signal nearly impossible to be observed.

Note that the variability in climate variables (air temperature, water vapor amount, and liquid and ice cloud water contents) of the rapidly rotating planet is much larger than that of the slowly rotating planet (Figure~\ref{fig_profiles}), but the variability in the transmission spectra is comparable between the slowly and rapidly rotating planets, as shown Figure~\ref{fig_spectra}. This is due to a ``saturation" effect: When the content of a climate variable, such as ice cloud water content, reaches a certain value, the optical depth will be far greater than one, so that the increment of the climate variable content will not significantly increase the atmospheric opacity, thus the transit atmospheric thickness remains almost the same. To explain this phenomenon, we varied the maximum value of ice cloud water profile for every grid along the terminator of a rapidly rotating planet. The results are shown in Figure~\ref{fig_along-lat}. When ice cloud water content exceeds 7 mg\,kg$^{-1}$, it becomes ``saturated", and the transit atmospheric thickness is almost the same as the original. Although the variability in climate variables of the rapidly rotating planet is rather large (Fig.~\ref{fig_profiles}), they are confined to low altitudes and are “saturated”, so they won’t contribute much to the variability. The amplitude of the variability depends mostly on the amount of ice cloud and water at high altitudes that are not ``saturated". 

\begin{figure}
    \includegraphics[width=1.0\textwidth]{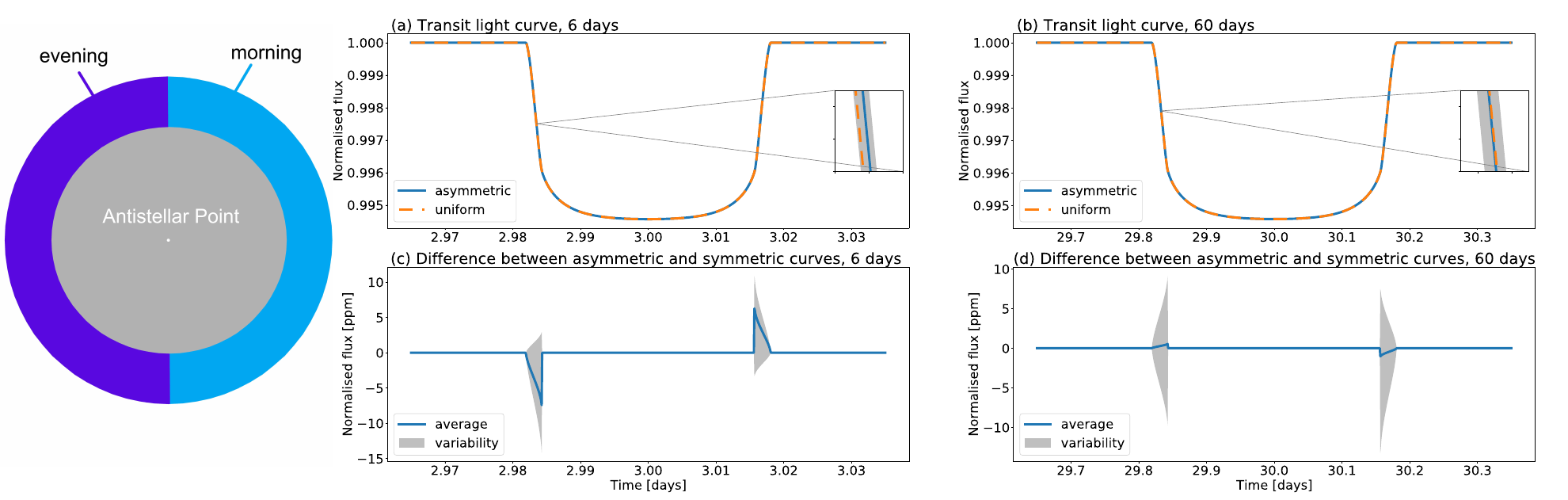}
    \caption{Light curves of primary transit for the real case (blue color) and an imagined symmetric planet (orange color). Panels (a) and (c) show the results for planets with a rotation period of 6 Earth days. Panels (b) and (d) show the results for planets with a rotation period of 60 Earth days. The lines are the mean values and the grey shading represents the variability. (c) and (d) are for the differences between the real case and the imaged symmetric planet. The light curves were calculated using the module PyTransit, a fast exoplanet transit model written in Python \citep{parviainen2015pytransit}, and the light curves are summed over 0.6 to 5 $\mu$m, the same as the spectral range of transmission spectra in Figure~\ref{fig_spectra}.}
    \label{fig_light_curves}
\end{figure}

We calculate the number of transits required to reach a given signal-to-noise (SNR) ratio, assuming that the overall SNR is in proportion to the square root of the number of transit events, and is also in proportion to the SNR for one transit, following the method used in \cite{lustig2019detectability}. Approximately, for a planet with a rotation period of 6 Earth days, it requires 19 transits to achieve 3$\sigma$ detection, and 53 transits to achieve 5$\sigma$ detection. The nominal lifetime of JWST is 5 yrs. If 30\% of the observation time is given to exoplanets, the possible maximum number of transits for an orbital period of 6~days is $\approx$90. Unlike the detection of molecular signatures, which are restricted to their absorption lines, the asymmetry signal exists in broad band-pass, so there is not much variation in the required number of transits for different wavelengths, except for several particular absorption lines. For a planet with a 60-day rotation period, it needs more than 300 transits to achieve 3$\sigma$ detection, which is impossible during the lifetime of JWST. In these calculations of the transit number, the variability was not considered; if it is included, the transit number would be at least doubled for a planet with a rotation period of 6 Earth days, since the 20 ppm variability is a bit larger than the output of PSG's noise calculator, which is about 15~ppm. The updated estimation is 55 transits to reach 3$\sigma$ detection, and 150 transits to reach 5$\sigma$ detection for the rapidly rotating case.

The asymmetry and variability are also reflected in the transit light curve, especially for the rapidly rotating case. Figure~\ref{fig_light_curves}(a) and (b) show the shape of the light curves of the real case and an imagined symmetric planet. In this study, we use the module of PyTransit \citep{parviainen2015pytransit} to calculate the transit light curves. The basic principle is summarized here: Considering an orbit without inclination, the normalized total flux $f$ during the transit is
\begin{equation}
f(d)=1-\frac{1}{\pi}\left[ (\frac{r}{R})^2\kappa_1+\kappa_2-\sqrt{\frac{d^2}{R^2} - \frac{(R^2+d^2-r^2)^2}{4R^4}} \right]
\end{equation}
where $\kappa_1=\arccos(\frac{r^2+d^2-R^2}{2rd}$), and $\kappa_2=\arccos(\frac{R^2+d^2-r^2}{2Rd})$ \citep{mandel2002analytic}. $R$ is the stellar radius, $r$ is the radius of the planet, and $d$ is the center-to-center distance between the star and the planet from front view, as shown in Figure 1(b) in \cite{mandel2002analytic}. We set $R$ the same as TRAPPIST-1, and assume there is no inclination of the orbit. During the ingress, $r=r_0+AT_{morning}$, and during the egress, $r=r_0+AT_{evening}$, where $r_0$ is the radius of TRAPPIST-1e, and $AT$ is the transit atmospheric thickness. The effect of limb darkening (i.e., the limb of the star appears darker than the center of the star) is considered using the method of \cite{mandel2002analytic} and  \cite{de2016combined}.

The asymmetry in the spatial patterns of clouds and water vapor results in a distortion of the light curve. The distortion for the rapidly rotating planet is larger than that for the slowly rotating planet, similar to the results in the transmission spectra. The distortion implies a late transit, but the timing offset is in the order of one second. The small magnitude of the offset, in addition to the confusing effect of the large variability (Figure~\ref{fig_light_curves}(c) and (d)), makes it very difficult to be detected in the observations of JWST.


\section{Summary and Discussion}\label{sec_summary}
In this study, we use PSG to calculate the transmission spectra of 1:1 tidally locked planets, based on the climatic states simulated using 3D global atmospheric circulation simulations. We focus on two important aspects that were not seriously considered in previous studies: the asymmetry between the morning and evening terminators and the variability of the transmission spectra. We find that there is significant asymmetry in the transmission spectra between the morning and evening terminators, especially for rapidly rotating planets. Ice clouds contribute the most to the asymmetry, due to the fact that they are at high levels, thus having a longer optical path. For rapidly rotating planets, the asymmetry is mainly caused by coupled Rossby-Kelvin waves and equatorial superrotation that transport water vapor and clouds to the evening terminator from the substellar region. Moreover, there is strong variability in the transmission spectra for both morning and evening terminators and for both rapidly and slowly rotating planets. This variability makes the detection of the asymmetry by JWST almost impossible.
      
In calculating the transit atmospheric thickness shown in Figures \ref{fig_spectra} and \ref{fig_light_curves}, we set both cases to have the same stellar radius, the radius of a 2600~K star. However, in reality, a star of 3700 K should have a radius three to four times of a 2600~K star. For example, the radius of TRAPPIST-1 with stellar temperature of 2516 $\pm$ 41~K is about 0.12 solar radius \citep{van2018stellar}, and the radius of Gliese 667 C, whose stellar temperature is 3700 $\pm$ 100~K \citep{anglada2012planetary}, is 0.42 solar radius. Considering the increment of stellar radius, the relative transit depth of the slowly rotating planet would be decreased by 10 to 20 times, so it is even more impossible to detect the asymmetry on a slowly rotating planet.

We only include N$_2$ and H$_2$O in this study. The atmospheric composition is simple and the planet is under a fixed stellar flux. If other species, such as O$_3$, which exists above the ice cloud deck, were considered in the simulations, the asymmetry might be even larger. If continents were included, the asymmetry may increase or decrease, depending on the location of the continents and the complex interactions among land, atmosphere, and clouds. In \cite{zhang2020does},  climatic states under different atmospheric pressures (from 0.5 to 4.0 bar) and stellar fluxes (from 1200 to 1800 W\,m$^{-2}$) had also been simulated. When using these climatic outputs to calculate the transmission spectra, we found that the magnitude of the asymmetry in all the experiments are between the rapidly rotating case and the slowly rotating case shown in section~\ref{sec_results}, and no clear trend of the transmission spectra as a function of the air pressure or stellar flux was found (figures not shown). We have also calculated the effect of oceanic heat transport on the magnitude of the asymmetry using the climatic outputs of coupled atmosphere--ocean general circulation experiments done in \cite{yang2019ocean}. We found that oceanic heat transport has a small effect (less than 3 ppm) on the asymmetry amplitude and the effect is negative under rapidly rotating orbits but positive under slowly rotating orbits (figures not shown). 

\cite{espinoza2021constraining} presented a semi-analytical framework to extract the transmission spectra of morning terminator and evening terminator directly from transit lightcurves, and showed that current missions like JWST and TESS are capable of detecting the asymmetry signal larger than 25 ppm. As shown in Figure 7 of their paper, the method is effective for hot Jupiters. For terrestrial planets, it will be more difficult, since the planet-to-star radius ratio and the asymmetry signal of terrestrial planets are smaller than those of hot Jupiters. Future larger telescopes with higher precisions will be more likely to detect the asymmetry signals between morning and evening terminators of habitable terrestrial planets.

Finally, we mention that the transmission spectra highly depend on the ice clouds at and around the terminators, but both convection and clouds in these models are parameterized based on the understandings of Earth. 
Because the scale of clouds is always smaller than the model grid sizes of GCMs, cloud water path and cloud fraction are parameterized. Different GCMs use different cloud parameterization schemes. Previous model intercomparisons for tidally locked habitable planets have shown that there are large differences in simulated cloud water path and cloud fraction between GCMs \citep{Yang_2019, fauchez2019trappist}. Many processes and parameters in the models can influence the simulation results. For example, the partition between liquid cloud and ice cloud is always parameterized based on air temperature in the microphysics modules. In ExoCAM, the fraction of cloud ice water in total condensate is set to be 100\% when air temperature is lower than -40$^{\circ}$C and to be 0\% when the air temperature is higher than -10$^{\circ}$C; between -40$^{\circ}$C and -10$^{\circ}$C, the fraction of cloud ice water is a linear function of the air temperature \citep{collins2004description}. In the model SAM, a similar linear function is used but the two temperature limits are set to be -20$^{\circ}$C and 0$^{\circ}$C \citep{khairoutdinov2003cloud}. For the latter, the cloud ice water path will be higher and the transmission thickness will be larger, if under the same condition. Another example is that the particle size of ice clouds is also parameterized in GCMs. In real world, the particle size as well as terminal velocity is determined by microphysical processes and should depend on air mass and planetary gravity \citep{loftus2021physics}, but this mechanism has not been included in present GCMs yet. In future studies, cloud resolving experiments with explicit convection and clouds are required to provide more confident conclusions.

\section*{Acknowledgments}

We are grateful to the help from Yixiao Zhang, Thaddeus D. Komacek, Xi Zhang, and Guo Chen. We thank Eric Wolf for the release of ExoCAM and thank NASA for the release of PSG. J.Y. acknowledges support from the National Natural Science Foundation of China (NSFC) under grant 41888101 and 42075046.

\bibliographystyle{frontiersinSCNS_ENG_HUMS}
\bibliography{frontiers}

\end{document}